\documentclass[a4paper]{jpconf}
\usepackage{graphicx}
\begin{document}
\title{Nuclear Excitations Described by\\
Randomly Selected Multiple Slater Determinants}

\author{S. Shinohara, H. Ohta, T. Nakatsukasa$^{*}$ and K. Yabana$^{*}$}

\address{Institule of Physics, University of Tsukuba, Tsukuba 305-8571, Japan\\
$^{*}$Center for Computational Sciences and Institute of Physics, University of Tsukuba, Tsukuba 305-8571, Japan}

\ead{sinohara@nucl.ph.tsukuba.ac.jp}

\begin{abstract}
We propose a new stochastic method to describe low-lying excited states of finite nuclei superposing multiple Slater determinants without assuming generator coordinates $a$ $priori$.
We examine accuracy of our method by using simple BKN interaction.
\end{abstract}.

\section{Introduction}
Self-consistent mean-field models using effective interactions (Skyrme, Gogny, etc) have been applied to description of ground-state properties of nuclei with great success for a wide mass region\cite{SHF1,SHF2}.
In addition, the generator coordinate method (GCM) is often utilized to take into account configuration mixing.
However, the GCM requires us to choose generator coordinates with physical intuition and the coordinate is restricted to one dimension in most cases.
Thus, it is difficult to describe many kinds of low-lying excitations simultaneously.
To overcome these difficulties, we attempt a new approach to select many Slater determinants.

\section{Method and results}
\begin{figure}[h]
\begin{minipage}{18pc}
\includegraphics[width=18pc]{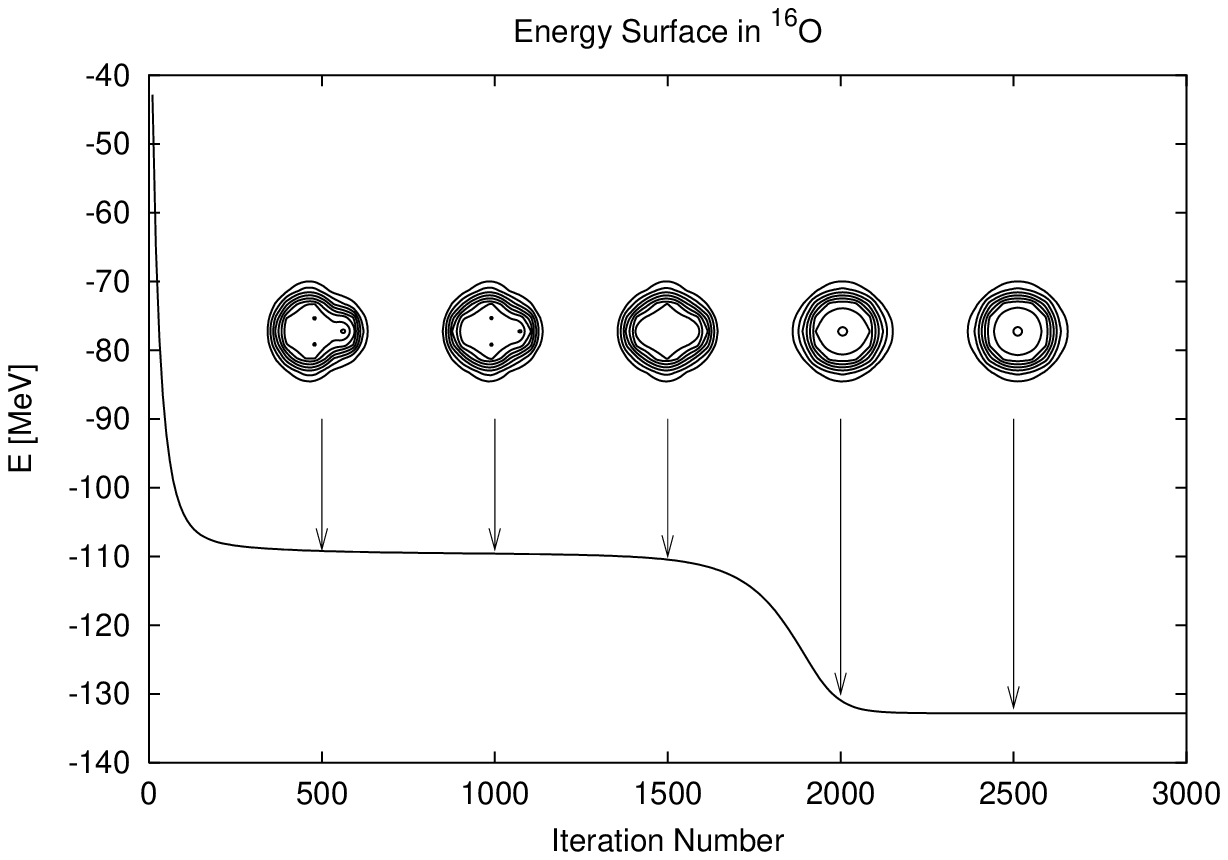}
\caption{\label{fig:itm}An example of the path generated by the imaginary-time calculation.
Energy as a function of the iteration and the density plots at every 500 iterations are shown.}
\end{minipage}\hspace{2pc}
\begin{minipage}{18pc}
\includegraphics[width=18pc]{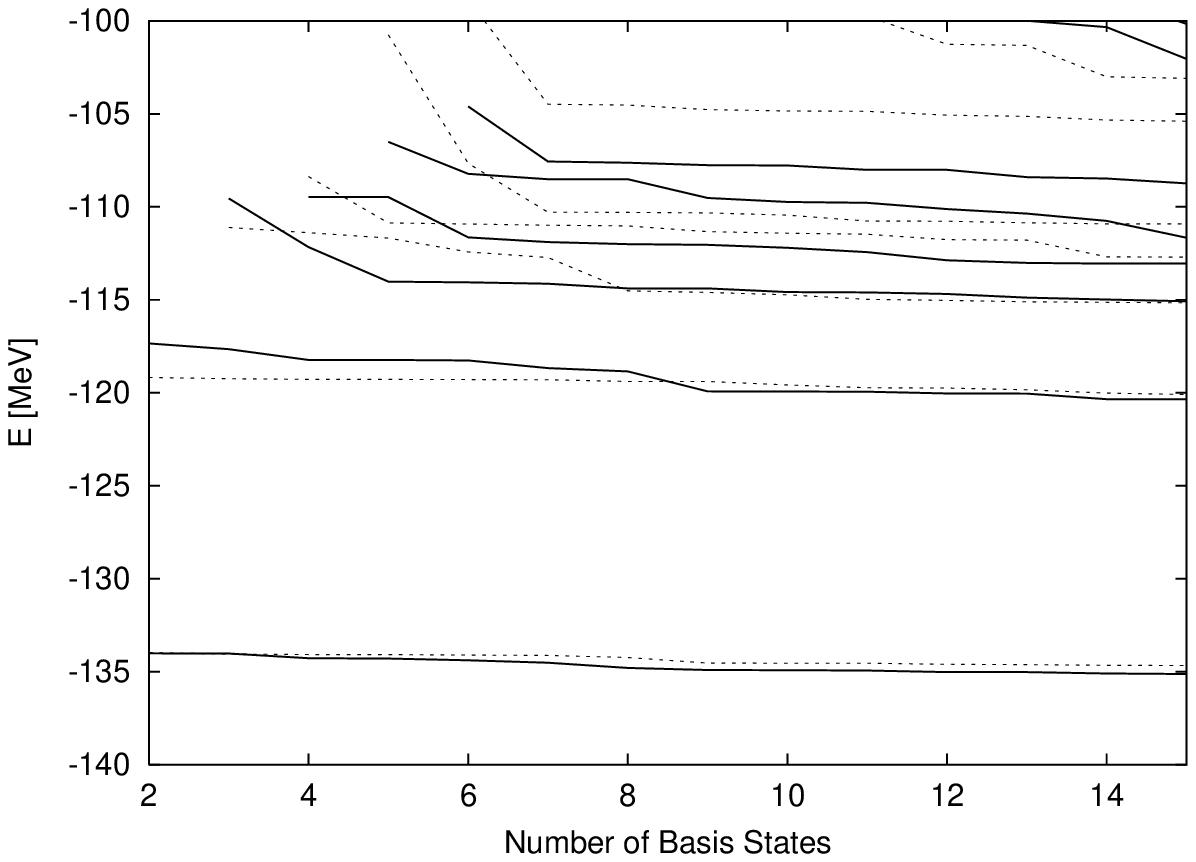}
\caption{\label{fig:energy}Energy of $J^{\pi}=0^{+}$ states in $^{16}{\rm O}$ as a function of number of basis states.
Solid and dotted lines indicate two independent calculations.}
\end{minipage} 
\end{figure}

We briefly discuss how to select Slater determinants $|\Phi^{s}\rangle\ (s=1,2,\cdots, N)$.
We use the imaginary-time method which is a standard procedure for solving self-consistent Hartree-Fock equations.
The energy expectation value of a Slater determinant is lowered as the imaginary time evolves and finally converges to the energy minimum state.
If the configuration approaches to a local minimum or a nearly stable
quasi-minimum (shoulder state), the state stays unchanged for a long period of time (Fig.\ref{fig:itm}).
We would like to select Slater determinants corresponding to these local minima and soft modes near the ground state.
For this purpose, we start imaginary-time calculations from many initial Slater determinants which are randomly generated, then pick up a state every 250 iterations of the imaginary-time step.
Since the state stays near local minima and the ground state for a long time, this prescription leads to a large probability of selecting states we want.
Repeating the same calculation with many initial states, we may select large number of Slater determinants without assuming generator coordinates.
We project each state on parity and angular momentum \cite{PPSHF}, then perform the configuration mixing calculation by solving the Hill-Wheeler-type equation \cite{RS,HW}.

It should be noted that, in order to avoid overcompleteness of selected set of basis, we exclude a state $|\Phi\rangle$ if this does not satisfy the condition,
\begin{eqnarray}
\ |\langle\Phi|
\left\{\!\!
\begin{array}{c}
1\\
P
\end{array}
\!\!\right\}
{\hat R}_{n}|\Phi^{s}\rangle|
<
0.7,
\end{eqnarray}
for any of selected states $\{|\Phi^{s}\rangle\}$.
${\hat R}_{n}$ $(n=1,2,\cdots,24)$ indicate special rotations corresponding to permutation of the axis ($x$, $y$, $z$).
$P$ is the parity-inversion operator.

Let us demonstrate that our method produces identical results for different initial sets of configurations generated by independent random numbers.
In Fig.\ref{fig:energy}, energy of $J^{\pi}=0^{+}$ states in $^{16}{\rm O}$ are shown as increasing number of Slater determinants.
Solid and dotted lines are independent calculations starting from different initial states.
We see that the calculated energies of lowest four levels are almost identical in these two calculations.
This indicates a qualitative convergence of the result with respect to the number of basis states.
Note that the obtained ground-state energy is lower than that of a single Slater determinant by about 2 MeV.

\section{Conclusion}
We proposed a new stochastic method for description of low-lying exited states of finite nuclei without assuming generator coordinates.
Each state is composed of superposition of multiple Slater determinants to include the correlation beyond the mean-field.
Basis states are generated by imaginary-time evolution with randomly-generated initial configurations.
The method is suitable for selecting local minima and soft modes.
We tested accuracy of our method by using simple BKN interaction.
Calculation with more realistic effective interaction is a future subject.

\clearpage
\end{document}